# Atomic Scattering Factor for a Spherical Wave and the Near Field Effects in X-ray Fluorescence Holography

Jianming Bai

Oak Ridge National Laboratory, Oak Ridge, TN 37831

Formula for calculating the atomic scattering factor for spherical x-ray waves is derived and used to solve the near field effects problem in X-ray Fluorescence Holography theory. A rigorous formalism to calculate the X-ray fluorescence hologram for given atomic distributions is then given so that quantitative structural information can be derived from XFH measurements.

PACS Number(s): 61:10.Dp, 61.10.Eq, 42.40.Kw, 32.80.Cy

**I. INTRODUCTION**

Atomic scattering factors (ASF) are used in numerous crystallographic calculations. The values of ASF for atoms and ions are tabulated in the International Tables for X-ray Crystallography. These ASFs are defined under the assumption that both the x-ray source and the detector are far away from the scatterers so that both the incident and scattered x-ray can be represented by plane waves. This assumption is valid for almost all kinds of x-ray scattering experiments until the emergence of x-ray fluorescence holography (XFH) in recent years.[1,2,3] XFH, because of its unique view point, is a new technique with great potential in exploring the local atomic orders. However, it needs further developments in two respects before becoming a routine method in material studies like XRD and EXAFS. First, because of the low signal/background ratio (typically $10^{-4}$ to $10^{-3}$) in XFH, it takes several days to collect data on a single hologram even with synchrotron radiation. The bottleneck is not the x-ray intensity but rather the detector speed. This problem will eventually be solved with developments of high-speed detectors. Secondly, the real space field intensity image reconstructed using the well-known Barton algorithm[4] is not a quantitative measurement of the atomic positions. One way to solve this problem is to use the nonlinear least-square-fitting algorithm to get accurate reconstruction of the atomic arrangement; this in turn requires a rigorous formalism to calculate the hologram. It is the task of this work to derive such formalism. In XFH, the fluorescence atoms inside the sample are used either as sources (in direct XFH) or as detectors (in inverse XFH, also called multiple energy x-ray holography, MEXF). The distance between the radiation source and scatterer, or the scatterer and the detector, is comparable to the size of the electron distribution of the scatterer. In this case, atomic scattering factors for a spherical wave should be used in the quantitative analysis for XFH. In all published work involving XFH calculation, a first order approximation has been used. This approximation assumes that the size of the core electron distribution of the scatterer is much smaller than the radius of the incident spherical wave front and is thus valid only for a point like scatterer.[5] In this paper, the formula to calculate the ASF for a spherical wave is derived and its dependence on the atomic radii of the scatterer, the source-scatterer or scatterer-detector distance and the wavelength has been examined. A detailed approach to calculate the spherical wave ASF from the electron density distribution or tabulated plane wave ASFs is developed. The only assumption is that the source (or detector) is outside of the electron distribution of the scatterer, which is always satisfied in XFH experiments. Given the similarity between the definitions of x-ray and electron ASF, the same scheme can also be used to calculate the spherical wave ASF for electrons in the high-energy regime, which can be used to correct the "small atom approximation" in photoelectron spectroscopy[6]. Another respect of the near field effects in XFH is originated from the vector property of the electromagnetic field. Using spherical wave ASF, these near field effects terms are also derived. With consideration of complete near field effects, rigorous expressions for calculating XFH and MXFH are given at the end.

**II, ATOMIC SCATTERING FACTOR FOR A SPHERICAL WAVE**

The X-ray fluorescence hologram is formed by the interference of the reference wave and the object waves. In direct XFH, the reference wave is the fluorescence radiation from the emitter atom and the object waves are the scattered fluorescence radiations from the surrounding atoms. In MEXF, the reference wave is the incident plane wave x-ray and the object waves are the scattered x-rays from the atoms surrounding the fluorescence emitter. The intensity of the fluorescence is a record of the interference pattern. In the frame of classical electrodynamics, the normalized x-ray fluorescence hologram is written as:

$$\chi(\mathbf{k}) = -r_e \sum_i \int d\mathbf{r} \frac{\rho(|\mathbf{r} - \mathbf{r}_i|)}{r} e^{i(kr - \mathbf{k} \cdot \mathbf{r})} + c.c., \qquad (1)$$

where $r_e$ is the classical electron radius and $\rho(\mathbf{r})$ is the electron charge density. The $\mathbf{r}$ is centered at the fluorescence emitter; $\mathbf{r}_i$'s are the centers of the scatterers and c.c. is the complex conjugation of the first term. To concentrate on the near field effects caused by the atomic scattering factors, we consider here only the scalar wave equation. The near field effects caused by the vector property of electromagnetic wave will be explored later. Let $\mathbf{u} = \mathbf{r} - \mathbf{r}_i$ and assume that $u \ll r_i$, we can have $r \approx r_i$ in the denominator and $r \approx r_i + \mathbf{r}_i \cdot \mathbf{u}/r_i$ in the phase term. This gives us the commonly used first approximation:

$$\chi(\mathbf{k}) \approx -r_e \sum_i \frac{f_i(s_i)}{r_i} e^{i(kr_i - \mathbf{k} \cdot \mathbf{r}_i)} + c.c., \qquad (2)$$

with $f_i(s_i) = \int d\mathbf{u} \rho_i(u) e^{-i(\mathbf{k} - \mathbf{k}_i) \cdot \mathbf{u}}$, $s_i = |\mathbf{k} - \mathbf{k}_i|$ and $\mathbf{k}_i = k\mathbf{r}_i / r_i$.

Here $f_i$ is just the ASF for plane wave x-rays. This approximation doesn't take the curvature of the spherical wave front into account and is valid only when $r_i$ is much larger than the radius of the scattering atom. This assumption is generally invalid for near neighbor atoms, and since the x-ray hologram is especially sensitive to near neighbor atoms, a more accurate formula is needed.

We define the spherical ASF as:

$$f_i^S(\theta_i, r_i) = r_i e^{-ikr_i + i\mathbf{k} \cdot \mathbf{r}_i} \int d\mathbf{r} \frac{\rho(|\mathbf{r} - \mathbf{r}_i|)}{r} e^{i(kr - \mathbf{k} \cdot \mathbf{r})} = r_i e^{-ikr_i} \int d\mathbf{u} \rho(u) \frac{e^{ikr}}{r} e^{-i\mathbf{k} \cdot \mathbf{u}}. \qquad (3)$$

By replacing the $f_i(s_i)$ in Eq.(2) with $f_i^S(\theta_i, r_i)$, one obtains the exact expression as given in Eq.(1).

The spherical wave term $e^{ikr}/r$ can be expanded with spherical harmonics:[7]

$$\frac{e^{ikr}}{r} = i4\pi k \sum_{l}(-1)^{l+m} j_l(ku_<)h_l^{(1)}(ku_>)\sum_{m}Y_l^m(\Omega_{r_i})Y_l^{-m}(\Omega_u), \tag{4}$$

and the plane wave term as:

$$e^{-i\mathbf{k}\cdot\mathbf{u}} = 4\pi\sum_{l',m'}(-i)^{l'}(-1)^{m'} j_{l'}(ku)Y_{l'}^{-m'}(\Omega_u)Y_{l'}^{m'}(\Omega_k). \tag{5}$$

Inserting Eq. (4) and (5) into Eq. (3), we have:

$$\begin{aligned} f_i^S(\theta_i, r_i) &= (4\pi)^2 r_i k i\, e^{-ikr_i} \sum_{l,m} i^l(-1)^m Y_l^m(\Omega_{r_i})Y_l^{-m}(\Omega_k) \int_0^\infty u^2 j_l(ku) j_l(ku_<) h_l^{(1)}(ku_>) \rho(u)\mathrm{d}u \\ &= 4\pi r_i k i\, e^{-ikr_i} \sum_l (2l+1) i^l P_l(\cos\theta_i)\left[ h_l^{(1)}(kr_i)\int_0^{r_i} j_l^2(ku)\rho(u)u^2 du + j_l(kr_i)\int_{r_i}^\infty j_l(ku)h_l^{(1)}(ku)\rho(u)u^2 du \right] \end{aligned}. \tag{6}$$

Where $\theta_i$ is the angle between $\mathbf{r}_i$ and $\mathbf{k}$. With a knowledge of the atomic electron density, Eq.(6) can be calculated for any $r_i$ value. However, in practice, it is more interesting to consider the situation when the source is outside of the electron distribution of the scattering atom. In this case, the second integration in Eq.(6) is zero. The plane wave ASF can be expanded with Legendre polynomial as:

$$f(|\mathbf{k}'-\mathbf{k}|) = 4\pi \sum_l (2l+1) P_l(\cos\theta) \int_0^\infty j_l^2(ku)\rho(u)u^2 du, \tag{7}$$

with $\theta$ defined as the angle between $\mathbf{k}$ and $\mathbf{k}'$, the Eq.(6) becomes:

$$f_i^S(\theta_i, r_i) = \frac{kr_i}{2} e^{-ikr_i} \sum_l (2l+1) i^{l+1} P_l(\cos\theta_i) h_l^{(1)}(kr_i) \int_0^\pi P_l(\cos\theta) f_i(2k\sin(\tfrac{\theta}{2}))\sin\theta d\theta . \tag{8}$$

One can easily verify Eq. (8) by assuming that $r_i$ is very large or that $\rho(u)$ is a delta function (so $f_i = z$), in both the cases $f_i^S(\theta_i, r_i)$ will degenerate into $f_i(s_i)$. The physical meaning of Eq. (8) is that the scattering power of an atom for a spherical wave can be represented by a weighted sum of plane wave ASF with the same wavelength in all directions.

However, when Eq. (8) is applied to a very small $r_i$ with tabulated plane wave ASF values, e.g. when $r_i$ equals 2.4825 Å, which is the nearest neighbor distance in a bcc iron crystal, the series in Eq. (8) is not convergent. This is because for free atoms the electron density continues beyond the nearest neighbor distance in their crystal form. From Figure 1 one can see that for an iron atom a small part of the electron distribution in 4s shell is outside of its first neighbor distance in the crystal. Therefore, the simple form of spherical wave ASF in Eq. (8) is not valid in this case since the first integral in Eq. (6) cannot be extended to infinity. Calculations show that Eq. (8) can be safely used for $r_i$ greater than twice the nearest neighbor distance for most elements. Spherical wave ASF can be calculated from Eq. (6) if the radial electron density is known. Even though the radial atomic electron densities can be very well calculated based on the shell model of atoms, their values are not conveniently available. It will be useful to have a formalism to calculate the spherical ASF with plane wave ASF values, which are experimentally measurable physical parameters. It is well known that the plane ASF is simply a three dimension Fourier image of the atomic electron distribution. For a spherical radial electron density model, we have:

$$f(s) = 4\pi \int_0^\infty \rho(r) \frac{\sin(sr)}{sr} r^2 dr, \tag{9}$$

and

$$\rho(r) = \frac{1}{2\pi^2 r} \int_0^\infty f(s)\sin(sr)s\,ds. \tag{10}$$

The tabulated ASF values in the *International Tables for Crystallography* are only available in the range of $s/4\pi$ from 0.0 to 6.0 Å$^{-1}$.[8] Evaluation of an accurate electron distribution requires a wider s range. However, what we need for evaluating the spherical ASF from Eq. (6) is the tail of the atomic electron density far from its nucleus, which is determined by the low s values of $f(s)$. We can calculate a partial electron density by constructing a partial ASF, $f_c(s)$, which is defined as:

$$f_c(s) = ae^{-bs^2} \text{ for } s \le s_c \text{ and } f_c(s) = f(s) \text{ for } s > s_c, \tag{11}$$

where $s_c$ is the cutoff value of $s$. The a and b are defined by setting the value and first derivative of $f_c(s)$ equal to those of $f(s)$ at the cutoff $s_c$:

$$a = f(s_c)e^{bs_c^2}, \quad b = -\frac{f'(s_c)}{2s_c f(s_c)}. \tag{12}$$

The $f_c(s)$ as defined corresponds only to the electron density close to the nucleus so that the partial electron density given by:

$$\rho_p(r) = \frac{1}{2\pi^2 r}\int_0^{s_c}(f(s)-f_c(s))\sin(sr)s\,ds = \frac{1}{2\pi^2 r}\int_0^\infty (f(s)-f_c(s))\sin(sr)s\,ds \tag{13}$$

will give the correct electron density for large r's. Figure 1 shows that the partial electron density calculated with a cutoff $s_c/4\pi=1.9$Å$^{-1}$ represents very well the entire 4s electron distribution in the iron atom. The partial ASF $f_c(s)$ represents only the electron distribution very close to the nucleus and can now be used in Eq. (8). The partial electron density which can be calculated with Eq. (13) with known plane wave ASF up to $s=s_c$ can be used in Eq. (6) to account for the contributions to the spherical ASF other than those from $f_c(s)$. Now the spherical wave ASF is given as:

$$f_i^S(\theta_i, r_i) = \frac{kr_i}{2}e^{-ikr_i}\sum_l (2l+1)i^{l+1}P_l(\cos\theta_i)c_l(k,r_i), \tag{14}$$

and

$$c_l(k,r_i) = h_l^{(1)}(kr_i)\int_0^\pi P_l(\cos\theta)f_c(2k\sin(\tfrac{\theta}{2}))\sin\theta\,d\theta \\ + 8\pi(h_l^{(1)}(kr_i)\int_0^{r_i} j_l^2(ku)\rho_p(u)u^2 du + j_l(kr_i)\int_{r_i}^\infty j_l(ku)h_l^{(1)}(ku)\rho_p(u)u^2 du) \tag{15}$$

Equation (15) can be used to calculate the spherical wave ASF for any physically meaningful $r_i$ values. The cutoff value $s_c/4\pi$ must be greater than $1/\lambda$ and less than the higher limit of the effective range of plane wave ASF value. Since 0 to 2.0 Å$^{-1}$ is the effective range of the widely used analytical representation of the plane wave ASF, the cutoff value of 1.9 Å$^{-1}$ used in the above example is a good choice for x-rays of energies less than 23 keV. In a crystal, the outmost electron distribution of an atom will be redistributed due to the neighbor atoms. Thus, the spherical symmetry is only approximately for the outmost electrons. Calculation shows that the contribution from the electron distribution outside $r_i$, which is given by the second integration in Eq. (15), is negligibly small (figure 2), thus the error caused by the spherical symmetry approximation should be small. It is well established that for plane wave ASF the x-ray reflection intensities are well represented by the free atom values of the form factors and are not very sensitive to the small redistributions of the electrons.[9]

As an example, the spherical ASF's were calculated for an iron atom with different $r_i$ (figure 3) and for different x-ray energies (figure 4). In all these calculations, except for the 40 kev curves, a cutoff $s_c/4\pi = 1.9$ Å$^{-1}$ and the four-Gaussian analytical representation of plane ASF by Doyle & Turner[10] were used. For the 40 keV curves, a cutoff $s_c/4\pi = 3.5$ Å$^{-1}$ and the five-Gaussian analytical representation of plane wave ASF by Waasmaier & Kirfel[11], which has an effective range from 0 to 6.0 Å$^{-1}$, were used. The calculation shows that the real parts (and the magnitude) of the spherical wave ASFs are about 10 to 20% less than the plane wave ASFs for the first neighbor scatterers around the forward scattering direction ($\theta_i = 0$) and approach the plane wave ASF values at high angles. This correction is mainly due to the curved wave front. The plane wave ASF will reach the electron number Z in forward scattering because in this direction the complete electron density distribution in the atom has the same phase. This will never happen for a spherical wave. The fact that the real part of the spherical wave correction vanishes at higher angles can be understood by looking at Eq. (9). The contribution to the ASF is mainly from electrons near the nucleus of the scattering atom for high $s$ since the function sin($sr$)/$sr$ acts like a δ-function for high $s$. The curved wave front correction is small for inner shell electrons. The imaginary part of the spherical ASF assumes a positive value about 10 to 20% of the atomic electron number and approaches a small negative constant at higher angles. This correction is a combination of the curved wave front effect and the $1/r$ dependence of the spherical wave amplitude. The $1/r$ weight in the electron distribution integration makes the apparent scatterer position closer to the source. Hence this contributes a negative phase shift, and this shift does not depend on the scattering angle. So the overall effect of the spherical wave ASF correction is that when the scatterer is between the source atom and the detector, it is 10 to 20% less in scattering power and apparently shifts away from the source. When the scatterer is on the opposite side of the source relative to the detector, it has an apparent position shift towards the source. The curved wave front correction is larger for shorter source-scatterer distance (figures 2 and 3) and higher x-ray energies (figure 4) as shown. Figure 2 shows the contribution to the spherical ASF from the electrons beyond $r_i$. The largest contribution is for the first neighbor scatterers and is less than 0.03 electrons for iron at 8 keV. The outer shell electron contribution is less for higher energies (not shown).

**III, NEAR FIELD EFFECTS DERIVED FROM VECTOR THEORY**

In addition to the near field effects caused by the spherical wave front, there are also near field effect terms caused by the vector property of the x-ray wave. In reference 5 these terms were considered under the plane wave approximation. Now with the spherical ASF, we can give a more accurate expression of these terms. In direct XFH, at the observation point **r** far away from the object, the total electric wave field is given by:[12]

$$\mathbf{E}(\mathbf{r}) = \nabla \times \nabla \times (g(r)\mathbf{p}) - \frac{r_e}{k^2} \nabla \times \nabla \times \int d\mathbf{r}' \left\{ g(|\mathbf{r}-\mathbf{r}'|) \rho(\mathbf{r}') \nabla' \times \nabla' \times [g(r')\mathbf{p}] \right\}. \tag{16}$$

Where g(r)=exp(ikr)/r and **p** is the electric dipole moment at **r**=0. Assuming $r \gg r'$, Eq. (16) can be simplified to:

$$\mathbf{E}(\mathbf{r}) = k^2 g(r)\mathbf{n} \times \left[\mathbf{p} - \frac{r_e}{k^2}\sum_i e^{-i\mathbf{k}\cdot\mathbf{r}_i}\nabla_i \times \nabla_i \times \left(\int d\mathbf{u} e^{-i\mathbf{k}\cdot\mathbf{u}}\rho(u)g(|\mathbf{r}_i - \mathbf{u}|)\mathbf{p}\right)\right] \times \mathbf{n}$$

$$= k^2 g(r)\mathbf{n} \times \left[\mathbf{p} - \frac{r_e}{k^2}\sum_i e^{-i\mathbf{k}\cdot\mathbf{r}_i}\nabla_i \times \nabla_i \times (\eta(r_i,\theta_i)\mathbf{p})\right] \times \mathbf{n} \quad (17)$$

Where $\mathbf{n}=\mathbf{r}/r$, $\eta(r_i,\theta_i)=g(r_i)f_i^S(\theta_i,r_i)$ and the sum is over all surrounding atoms. This leads to an expression for the hologram:

$$\chi(\mathbf{k}) = \sum_i \chi_i(\mathbf{k}) \text{ with } \chi_i(k) = -\frac{r_e}{k^2}e^{-i\mathbf{k}\cdot\mathbf{r}_i}\nabla_i \times \nabla_i \times (\eta(r_i,\theta_i)\mathbf{p})\cdot\mathbf{p}_n/\mathbf{p}\cdot\mathbf{p}_n + c.c. \quad (18)$$

Where $\mathbf{p}_n = \mathbf{n}\times\mathbf{p}\times\mathbf{n} = \mathbf{p} - (\mathbf{n}\cdot\mathbf{p})\mathbf{n}$. After some tedious but straightforward vector algebra, we have:

$$\chi_i(\mathbf{k}) = -\frac{r_e}{k^2}e^{-i\mathbf{k}\cdot\mathbf{r}_i}\{-\mathbf{p}\cdot\mathbf{p}_n\nabla_i^2\eta(r_i,\theta_i) + (\mathbf{p}\cdot\hat{\mathbf{r}}_i)(\mathbf{p}_n\cdot\hat{\mathbf{r}}_i)\frac{\partial^2}{\partial r_i^2}\eta(r_i,\theta_i)$$

$$+[(\mathbf{p}\cdot\hat{\mathbf{r}}_i)(\mathbf{p}_n\cdot\hat{\theta}_i) + (\mathbf{p}_n\cdot\hat{\mathbf{r}}_i)(\mathbf{p}\cdot\hat{\theta}_i)]\frac{\partial}{\partial r_i}(\frac{1}{r_i}\frac{\partial}{\partial\theta_i})\eta(r_i,\theta_i)$$

$$+(\mathbf{p}\cdot\hat{\theta}_i)(\mathbf{p}_n\cdot\hat{\theta}_i)\frac{1}{r_i}(\frac{\partial}{\partial r_i} + \frac{1}{r_i}\frac{\partial^2}{\partial\theta_i^2})\eta(r_i,\theta_i) \quad (19)$$

$$+(\mathbf{p}\cdot\hat{\varphi}_i)(\mathbf{p}_n\cdot\hat{\varphi}_i)\frac{1}{r_i}(\frac{\partial}{\partial r_i} + \frac{1}{r_i\tan(\theta_i)}\frac{\partial}{\partial\theta_i})\eta(r_i,\theta_i)\}/\mathbf{p}\cdot\mathbf{p}_n + c.c.$$

For direct XFH, the electric dipole moment $\mathbf{p}$ is averaged in $4\pi$ solid angles since the fluorescence radiation is unpolarized. So we now have:

$$\chi_i(\mathbf{k}) = -r_e e^{-i\mathbf{k}\cdot\mathbf{r}_i}\{\eta(r_i,\theta_i)(1+\cos^2\theta_i)/2$$

$$+\frac{1}{2k^2}[\frac{3\cos^2\theta_i - 1}{r_i}\frac{\partial}{\partial r_i}\eta(r_i,\theta_i) + \frac{\cos(2\theta_i)}{r_i^2}\frac{\partial^2}{\partial\theta_i^2}\eta(r_i,\theta_i) \quad (20)$$

$$+\frac{\sin(2\theta_i)}{r_i}\frac{\partial^2}{\partial r_i\partial\theta_i}\eta(r_i,\theta_i) + \frac{1-3\sin^2\theta_i}{r_i^2\tan\theta_i}\frac{\partial}{\partial\theta_i}\eta(r_i,\theta_i)]\} + c.c.$$

In deriving equation (20), please note that the function $\eta(r_i,\theta_i)$ satisfies the Helmholtz differential equation: $\nabla_i^2\eta + k^2\eta = 0$. The derivatives of $\eta(r_i,\theta_i)$ can be calculated using Eq. (8). For example:

$$\frac{\partial\eta(r_i,\theta_i)}{\partial\theta_i} = \frac{k}{2}\sum_l(2l+1)i^{l+1}P_l^1(\cos\theta_i)h_l^{(1)}(kr_i)\int_0^\pi P_l(\cos\theta)f_i(2k\sin(\tfrac{\theta}{2}))\sin\theta d\theta.$$

Where $P_l^1(\cos\theta_i)$ is associated Legendre polynomial of order $l$ and degree 1.

For MXFH, assume the incident x-ray wave is polarized and the $\mathbf{E}$ vector is perpendicular to $\mathbf{k}$ and surface normal of the sample, as in the case of synchrotron radiation, we have:

$$\chi_i(\mathbf{k}) = -r_e e^{-i\mathbf{k}\cdot\mathbf{r}_i} \{\eta(r_i,\theta_i)(1-\sin^2\theta_i\cos^2\varphi_i)$$

$$+\frac{1}{k^2}[\frac{1-3\sin^2\theta_i\cos^2\varphi_i}{r_i}\frac{\partial}{\partial r_i}\eta(r_i,\theta_i) + \frac{\cos(2\theta_i)\cos^2\varphi_i}{r_i^2}\frac{\partial^2}{\partial\theta_i^2}\eta(r_i,\theta_i) \quad\quad (21)$$

$$+\frac{\sin(2\theta_i)\cos^2\varphi_i}{r_i}\frac{\partial^2}{\partial r_i\partial\theta_i}\eta(r_i,\theta_i) + \frac{1-\cos^2\varphi_i(1+3\sin^2\theta_i)}{r_i^2\tan\theta_i}\frac{\partial}{\partial\theta_i}\eta(r_i,\theta_i)]\} + c.c.$$

If the incident beam is unpolarized, Eq. (21) needs to be averaged over $\varphi_i$ and will be the same as Eq. (20). In Eq. (20) and (21), the $\chi_i(\mathbf{k})$'s dependence on the direction of $\mathbf{k}$ is through $\theta_i$, which is the angle between $\mathbf{k}$ and $\mathbf{r}_i$; and $\varphi_i$, the angle between $\mathbf{p}$ and $\mathbf{r}_i - (\mathbf{n}\cdot\mathbf{r}_i)\mathbf{n}$. They can be expressed in terms of the spherical coordinates of $\mathbf{k}$ ($\theta, \varphi$) and $\mathbf{r}_i$ ($\theta'_i, \varphi'_i$) in a coordinator system fixed on the sample:

$$\cos\theta_i = \cos\theta\cos\theta'_i + \sin\theta\sin\theta'_i\cos(\varphi-\varphi'_i)$$
$$\cos\varphi_i = \sin(\varphi-\varphi'_i)\sin\theta'_i/\sin\theta_i \quad\quad (22)$$

Combing equations (18), (20), (21) and (22), the x-ray fluorescence holograms can be calculated with consideration of complete near field effects. Fig. 5 shows a calculated $\chi_i(\theta,\varphi=0)$ curve for a single emitter-scatterer pair of iron atoms. The fluorescence emitter atom is at origin while the scatterer is at 2.4825 Å along the x axis (e.g. $\theta'_i = \varphi'_i = 0$). The calculation shows that at x-ray energy of 6.4 keV, the corrections due two the two kinds of the near field effects are comparable. At x-ray energy of 20 keV, the correction due to the curved wave front is larger while the correction due to the near field effect caused by the vector field properties is smaller. This is because of the $1/k^2$ dependence in Eq. (20).

A generalized atomic scattering factor can be defined as function of $\theta_i$ and $r_i$ for XFH or MXFH with unpolarized incident beam:

$$f_g(r_i,\theta_i) = f_i^S(r_i,\theta_i)(1+\cos^2\theta_i)/2$$

$$+\frac{e^{-ikr_i}}{2k^2}[(3\cos^2\theta_i - 1)\frac{\partial}{\partial r_i}\eta(r_i,\theta_i) + \frac{\cos(2\theta_i)}{r_i}\frac{\partial^2}{\partial\theta_i^2}\eta(r_i,\theta_i) \quad\quad (23)$$

$$+\sin(2\theta_i)\frac{\partial^2}{\partial r_i\partial\theta_i}\eta(r_i,\theta_i) + \frac{1-3\sin^2\theta_i}{r_i\tan\theta_i}\frac{\partial}{\partial\theta_i}\eta(r_i,\theta_i)]$$

With this generalized atomic scattering factor, the simple scalar form as of equation (2) can be used to calculate the hologram in XFH or MXFH with unpolarized incident beam. The near field effect caused by curved wave front is contained in the spherical wave ASF $f_i^S(r_i,\theta_i)$ while the near field effect caused by the vector property is presented in the four terms in the squared parentheses. Fig. 6 shows the real and imaginary part of the generalized ASF for iron atom at 6.4 keV as function of $\theta_i$ with $r_i$=2.4825 Å. Since the generalized ASF is a slow varying function of $r_i$ and $\theta_i$, they can be stored in a two dimension array at increment $r_i$ and $\theta_i$ values. When calculating the x-ray hologram, the values of the generalized ASF at any $r_i$ and $\theta_i$ values can be retrieved by interpolation. For MXFH with polarized incident beam (Eq. 21), the generalized ASF can be divided into two parts, $f_g = f_g^1(r_i,\theta_i) + f_g^2(r_i,\theta_i)\cos^2\varphi_i$. In this paper we consider only the Thomson scattering. When x-ray energy is close to the absorption edge, the complex anomalous scattering factor should be added to the generalized ASF.

## IV. CONCLUSION

In conclusion, the formalism for calculating the spherical wave ASFs from the radial electron distribution function or the tabulated plane wave ASFs is derived. The example calculation shows that the curved wave front correction to ASFs can be as large as 20% of the plane wave ASF, depends on the source-scatterer distance and x-ray energy. This correction should be considered in all XFH or MXFH calculation. On the other hand, the near field effects due to the vector property of the electromagnetic field is comparable to curved wave front correction for low x-ray energies and must also be included for most of XFH calculation. With both kinds of near field effects considered, the rigorous expressions for both XFH and MXFH are given. These expressions are necessary parts in the developing of new reconstruction method based on least- square-fitting algorithms.

There are other approximations in XFH theory, i.e. the single scattering approximation and the point dipole moment source approximation for the fluorescence atoms. Unlike photoelectron holography, in XFH the multiple scattering effects are generally negligible due to the small cross section of x-ray scattering. On the hand, since only the inner shell fluorescence is used for XFH experiment, the size of the radiation source is also negligible. For example, the radius of 1s shell in Fe atom is 0.03 Å, which is only about 1.3% of the nearest neighbor distance in BCC iron crystal and much smaller than the wavelength of iron kα line. Therefore, the error caused by the point dipole moment field approximation should be one magnitude smaller than the corrections considered in this paper. Moreover, because of the spherical symmetry of the s-shell electron distribution, the dipole moment size effect should be isotropic and hardly detectable in XFH measurements.

## ACKNOWLEDGEMENTS

The author thanks Camden Hubbard, Gene Ice and Cullie Sparks for reviewing the manuscript and helpful discussions. Research sponsored by the Assistant Secretary for Energy Efficiency and Renewable Energy, Office of Transportation Technologies, as part of High Temperature Materials Laboratory User Program, Oak Ridge National Laboratory, managed by UT-Battelle, LLC, for the U.S. Dept. of Energy under contract DE-AC05-00OR22725.

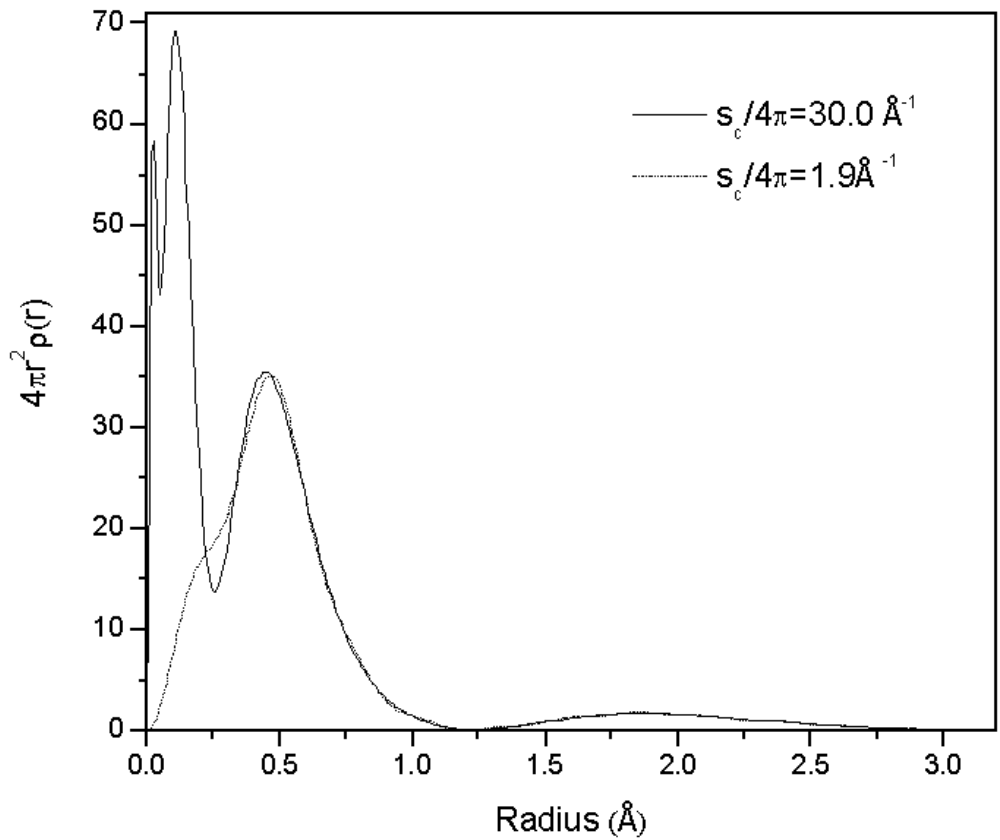

Fig. 1 Partial electron distributions of atom iron calculated with equation (13). The plane wave ASF for iron used here is based on an analytical interpolation of one-electron wave functions built to approximate the solution to Hartree-Fock equations and is valid for entire range of $s$.[13] For a cutoff $s$ of 30 Å$^{-1}$, the partial electron distribution calculated (solid line) is practically the full electron distribution of atom iron.

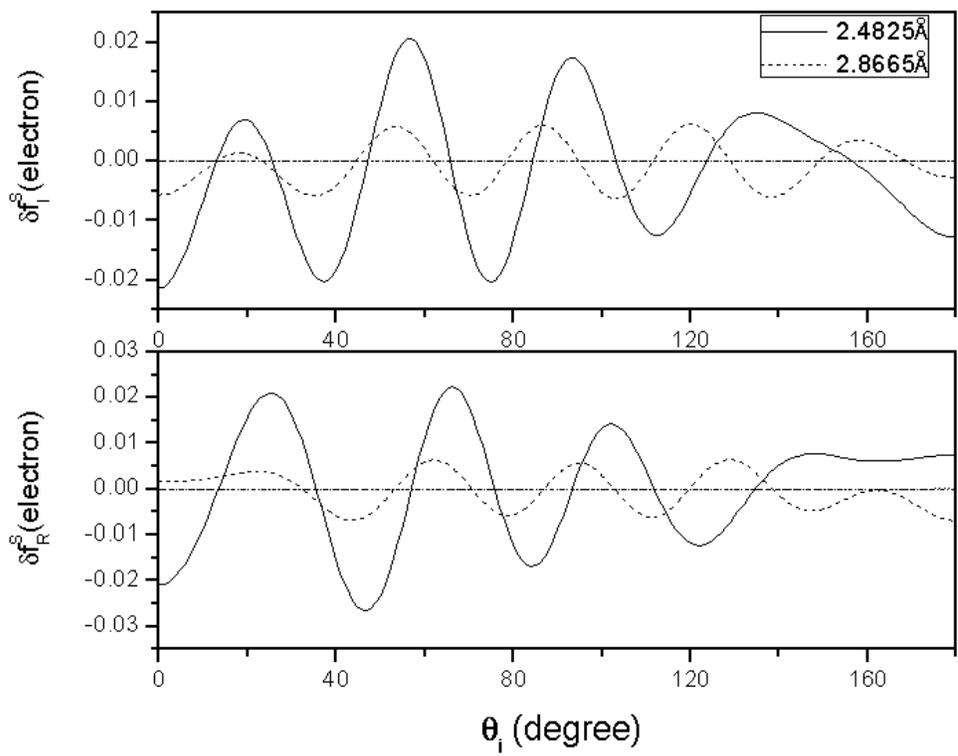

Fig. 2, The imaginary (top) and real part (bottom) of the contributions to the spherical wave ASF for iron from the electron distributions beyond of its first and second neighbor distance. The x-ray energy is 8 keV.

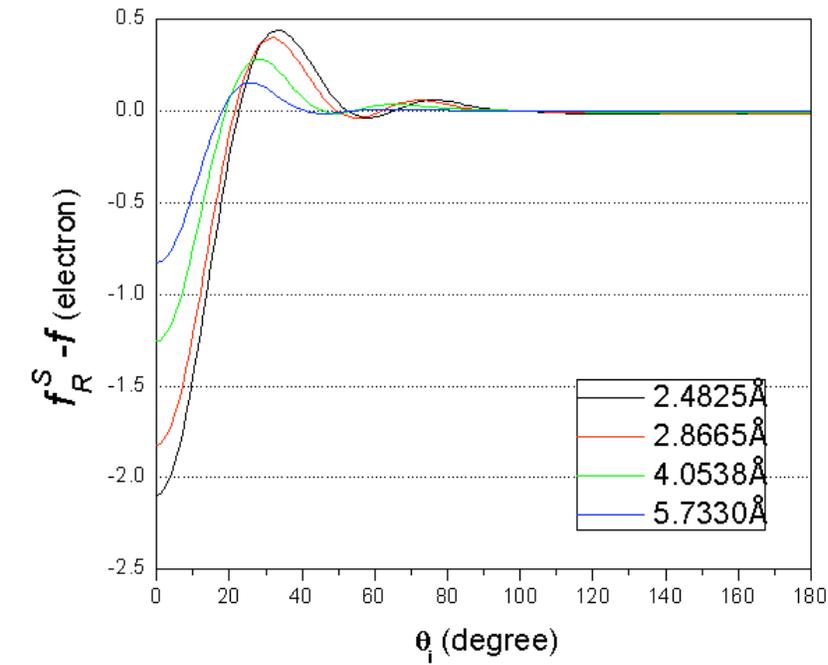

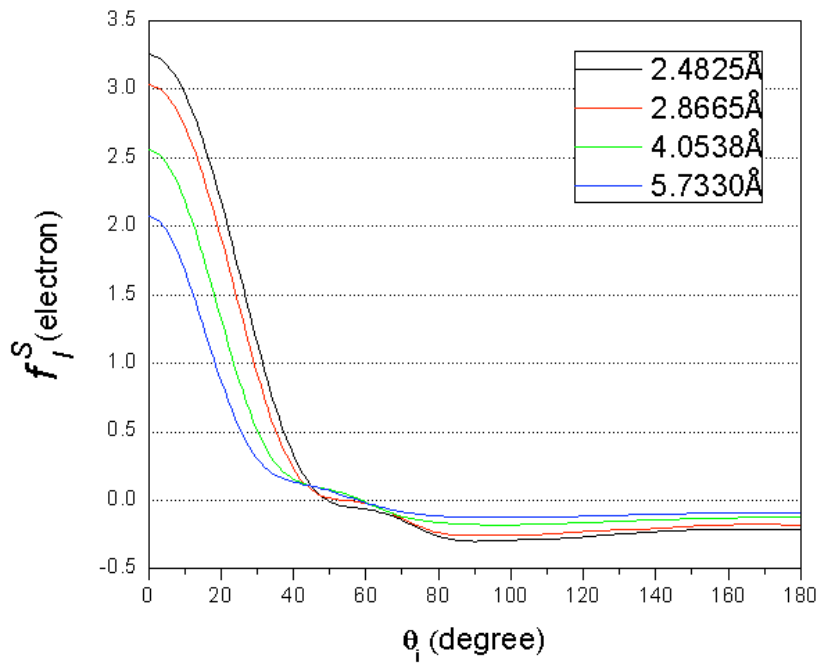

Fig. 3, The difference between the real part of spherical wave ASF and the plane wave ASF (top) and the imaginary part of the spherical wave ASF (bottom) for iron calculated with $r_i$ equal to its first, second, third and fourth neighbor distance at 8 keV.

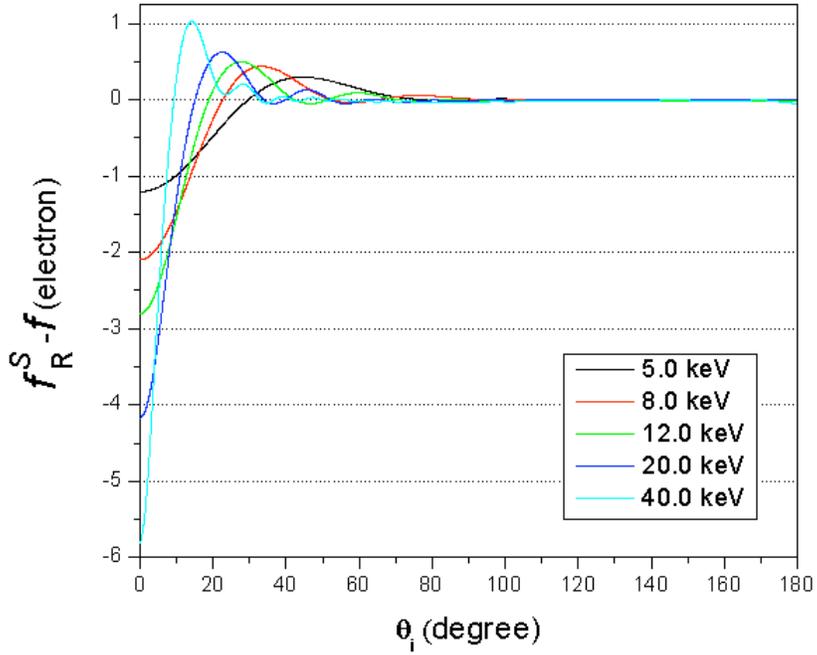

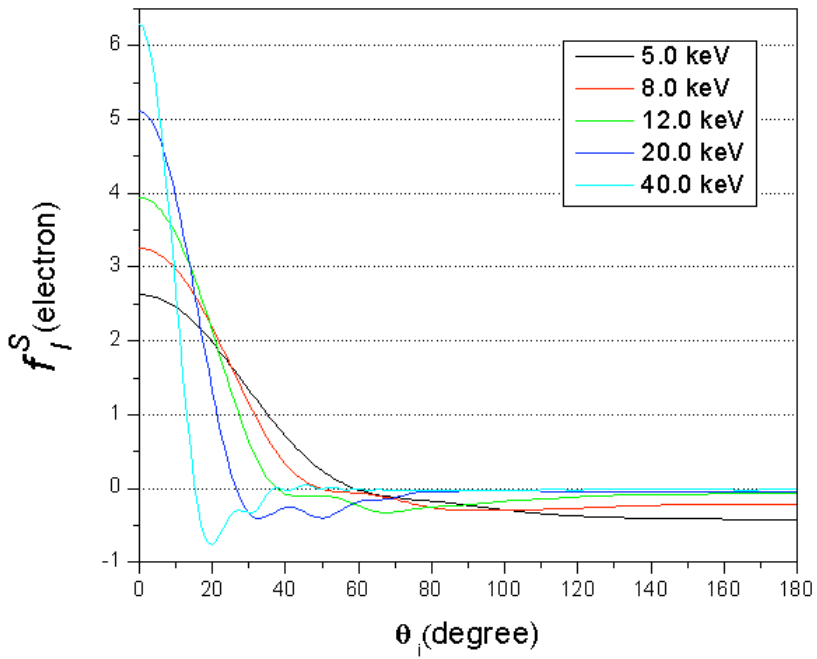

Fig. 4, X-ray energy dependence the ASF for a spherical wave. Top: real part of the spherical ASF minus the plane ASF. Bottom: imaginary part.

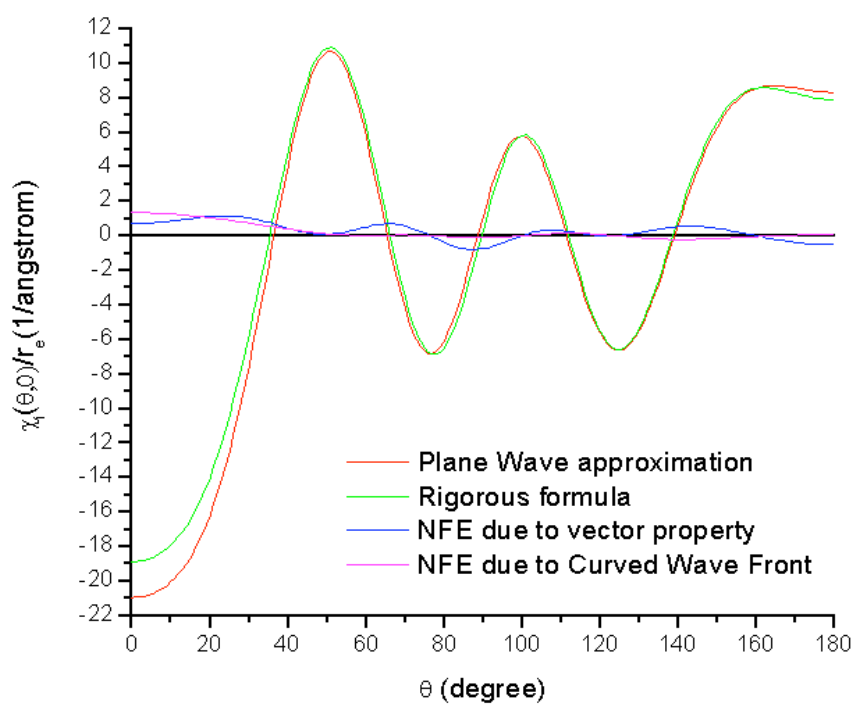

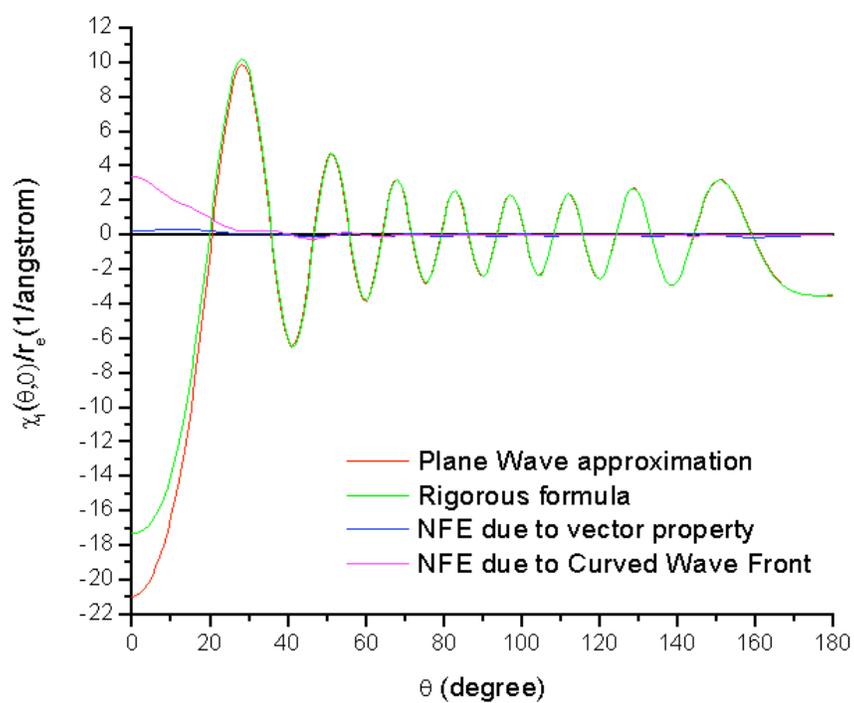

Fig. 5, X-ray fluorescence hologram curves for a single pair of iron atoms separated by 2.4825 Å. Top: E=6.4 keV. Bottom: E=20 keV.

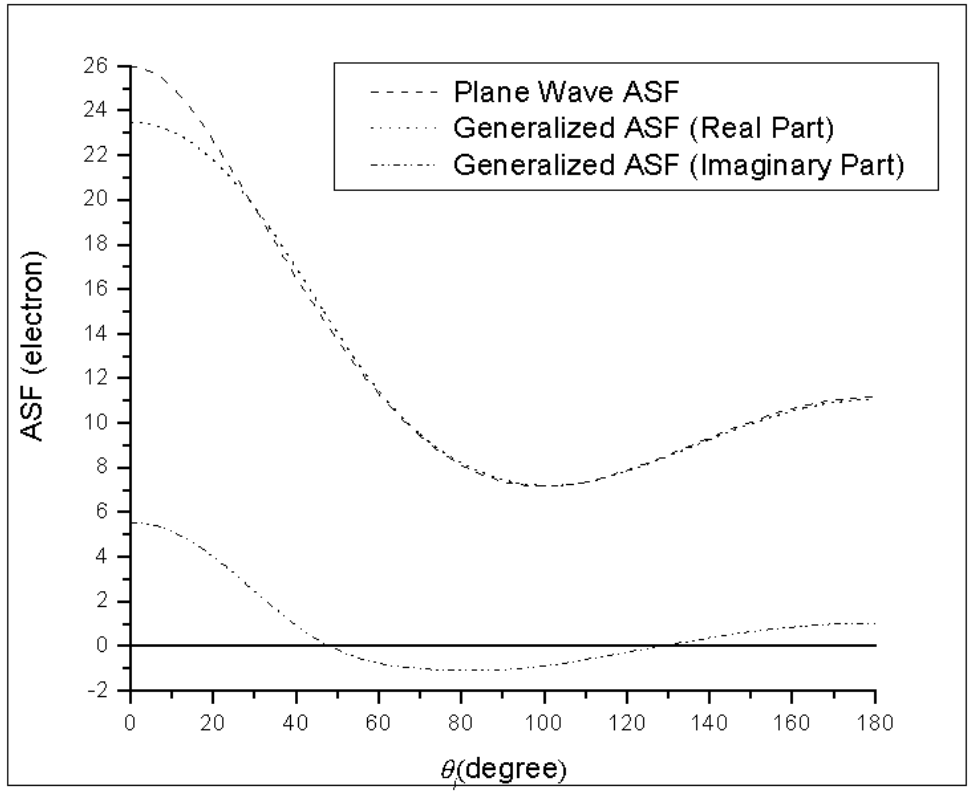

Fig. 6, The real and imaginary part of the generalized ASF with $r_i$ = 2.4825Å for iron atom and x-ray energy of 6.4 keV (Fe kα-lines). For comparison, the plane wave ASF multiplied by the polarization factor $(1+\cos^2\theta_i)/2$ is also plotted.